\documentclass[5p,sort&compress]{elsarticle}

\usepackage[T1]{fontenc}
\usepackage[utf8]{inputenc}
\usepackage{amsmath}
\usepackage{amssymb}
\usepackage{mathrsfs}

\journal{Mechanics Research Communications: https://doi.org/10.1016/j.mechrescom.2018.08.001}

\newcommand{\p}{\partial}

\begin{document}

  \begin{frontmatter}

    \title{Thermal effects on nonlinear acceleration waves \\
    	in the Biot theory of porous media}

    \author[durham]{Brian Straughan}
    \ead{brian.straughan@durham.ac.uk}

    \author[unisa]{Vincenzo Tibullo\corref{cor1}}
    \ead{vtibullo@unisa.it}

    \cortext[cor1]{Corresponding author}

    \address[durham]{Durham University, Department of Mathematical Sciences}
    \address[unisa]{Università di Salerno, Dipartimento di Matematica}

    \begin{abstract}
    	We generalize a theory of Biot for a porous solid based on nonlinear elasticity theory to incorporate temperature effects.
		Acceleration waves are studied in detail in the fully nonlinear theory.
		The wavespeeds are found explicitly and the amplitudes are then determined.
		The possibility of shock formation is discussed.
    \end{abstract}


  \end{frontmatter}

\section{Introduction}

	The topic of wave propagation in porous and acoustic media is one of great interest in the current research literature, see e.g.\ \citet{Biot1962}, \citet{Brunnhuber2016}, \citet{Christov2016a}, \citet{Christov2010}, \citet{Christov2016b}, \citeauthor{Ciarletta2006}  \citep{Ciarletta2006, Ciarletta2007a, Ciarletta2007b}, \citeauthor{Jordan2004} \citep{Jordan2004, Jordan2005a, Jordan2006, Jordan2009, Jordan2013, Jordan2016}, \citet{Jordan2005b}, \citet{Jordan2012}, \citet{Jordan2017, Jordan2018}, \citet{Paoletti2012}, \citeauthor{Rossmanith2016a} \citep{Rossmanith2016a, Rossmanith2016b}, \citet{Wei2013}. 

	In a recent paper, \citet{Ciarletta2018}, we developed a fully nonlinear acceleration wave analysis for an isothermal theory of porous media which incorporates finite deformation in nonlinear elasticity, this theory having been proposed by \citet{Biot1972}.
	This work extends previous work by \citet{Jordan2005a} and by \citet{Ciarletta2006} who studied nonlinear acoustic waves in a porous medium when the elastic skeleton is rigid.
	The aim of the present article is to incorporate temperature effects into the Biot model and then extend the analysis of \citet{Ciarletta2018} to the non-isothermal situation.
	We emphasize that this is not a trivial extension since we find the elastic wave, the pressure wave, and the temperature wave are intrinsically coupled.
	
	When the skeleton in a porous body is allowed to deform it is not trivial to analyse nonlinear wave motion.
	One way to achieve this has employed a theory of a mixture of a fluid and an elastic solid, see e.g.\ \citet{DeBoer1995}.
	A second way is to include a distribution of voids in an elastic body, see e.g.\ \citet{Iesan2004}, \citeauthor{Ciarletta2007a} \citep{Ciarletta2007a, Ciarletta2007b}.
	A third way is via the Biot pressure function theory, see \citet{Biot1972}, \citet{Ciarletta2018}.
	A comparison of wave motion in the last two mentioned theories is given in chapter 4 of \citet{Straughan2017}.
	It is worth pointing out that \citet{Biot1972} is critical of employing a mixture theory approach, and \citet{Chen1984} is likewise critical of using acceleration waves in mixture theories.

\section{Nonlinear thermoelastic theory of porous media}

	We use standard indicial notation in conjunction with the Einstein summation convention throughout.
	We denote points in the reference configuration by $X_A$ and these are transformed into the current configuration by the mapping
\begin{equation}
	x_i = x_i(X_A, t).
\end{equation}
	The deformation gradient $F_{iA}$ and displacement vector $u_i$ are defined by
\begin{align}
	F_{iA} &= \frac{\p x_i}{\p X_A}, \\
	u_i    &= x_i-X_i.
\end{align}
	
	We commence with the momentum equation, cf.\ \citet{Biot1972}, \citet{Ciarletta2018},
\begin{equation}
	\rho\ddot{x}_i = \frac{\p\Pi_{Ai}}{\p X_A}+\rho f_i,
	\label{eq:motion}
\end{equation}
where $\rho$ is the density (in the reference configuration), $\Pi_{Ai}$ is the Piola-Kirchhoff stress tensor, and $f_i$ denotes the body force. 
	A superposed dot denotes differentiation with respect to time.
	
	For a thermoelastic body the balance of energy equation may be written as, see e.g.\ \citet{Straughan2017}, p.53,
\begin{equation}
	\rho\theta\dot\eta = -\frac{\p q_A}{\p X_A}+\rho r,
	\label{eq:energy}
\end{equation}
where $\theta(\mathbf{X}, t)$ is the temperature, $\eta$ is the specific entropy, $q_A$ is the heat flux, and $r$ denotes an externally supplied heat source.
	In terms of the Helmholtz free energy function $\psi$ the entropy is given by
\begin{equation}
	\eta = -\frac{\p\psi}{\p\theta}.
\end{equation}
	
	The \citet{Biot1972} theory involves the pressure, $p(\mathbf{X}, t)$, in the pores in the elastic body.
	This theory employs a conservation law for the pressure function which may be written as
\begin{equation}
	\frac{\p m}{\p t} = \frac{\p J_A}{\p X_A}.
	\label{eq:m-motion}
\end{equation}
	In equation \eqref{eq:m-motion} $J_A$ is a flux term and both $J_A$ and $m$ depend on $F_{iA}$ and upon a function $\phi$ which is a nonlinear function of the pressure $p$.

	In this work we propose a thermoelastic theory for porous media based upon equations \eqref{eq:motion}, \eqref{eq:energy} and \eqref{eq:m-motion} and we propose that $\psi$ and $m$ depend on the variables
\begin{equation}
	\begin{aligned}
		\psi &= \psi(F_{iA}, p, \theta, X_A) \\
		 m   &= m   (F_{iA}, p, \theta, X_A)
	\end{aligned}
	\label{eq:fields}
\end{equation}
while the fluxes $J_A$ and $q_A$ have the functional dependence
\begin{equation}
	\begin{aligned}
		q_A &= q_A(F_{iB}, p, p_{,B}, \theta, \theta_{,B}, X_B) \\
		J_A &= J_A(F_{iB}, p, p_{,B}, \theta, \theta_{,B}, X_B).
	\end{aligned}
	\label{eq:fluxes}
\end{equation}
	
	Without loss of generality we now set the body force $f_i$ and heat supply $r$ to be zero.

\section{Nonlinear acceleration waves}

	The governing system of equations is \eqref{eq:motion}, \eqref{eq:energy} and \eqref{eq:m-motion} and we define an acceleration wave for a solution to this system to be a singular surface $\mathscr{S}$ across which $\ddot{x}_i$, $\dot{x}_{i,A}$, $x_{i,AB}$, $\ddot{p}$, $\dot{p}_{,A}$, $p_{,AB}$, $\ddot\theta$, $\dot\theta_{,A}$ and $\theta_{,AB}$ and their higher derivatives suffer a finite discontinuity, but $x_i$, $p$, $\theta$ are continuously differentiable throughout $\mathbb{R}^3$ for $t\in[0,\mathcal{T}]$ for some time interval.
	
	Nonlinear acceleration wave analysis is well known, see e.g.\ \citet{Chen1973}, and so we give minimal details of the calculations. 
	One employs the constitutive theory \eqref{eq:fields} and \eqref{eq:fluxes} in equations \eqref{eq:motion}, \eqref{eq:energy} and \eqref{eq:m-motion} and then we take the jumps of the resulting equations. 
	We find that
\begin{equation}
	[\ddot{x}_i] = \frac{\p^2\psi}{\p F_{jB}\p F_{iA}}[x_{j,AB}],
	\label{eq:x-jump}
\end{equation}
where we have used the fact that $\Pi_{Ai} = \rho\p\psi/\p F_{iA}$, and we further obtain
\begin{align}
	& \frac{\p m}{\p F_{iA}}[\dot{F}_{iA}] = 
		\frac{\p J_A}{\p F_{jB}}[F_{jB,A}]+
		\frac{\p J_A}{\p p_{,B}}[p_{,BA}]+
		\frac{\p J_A}{\p \theta_{,B}}[\theta_{,BA}],
		\label{eq:m-jump} \\
	& \rho\theta\frac{\p\eta}{\p F_{iA}}[\dot{F}_{iA}] = 
		-\frac{\p q_A}{\p F_{iK}}[F_{iK,A}]-
		 \frac{\p q_A}{\p\theta_{,K}}[\theta_{,KA}]-
		 \frac{\p q_A}{\p p_{,K}}[p_{,KA}],
		 \label{eq:eta-jump}
\end{align}
where $[\,\cdot\,]$ denotes the jump of a quantity, eg.\ $[f]=f^- - f^+$.

	Define now the amplitudes $a_i$, $b$ and $c$ by
\begin{equation}
	a_i(t) = [\ddot{u}_i],\qquad b(t) = [\ddot{p}],\qquad c(t) = [\ddot\theta].
\end{equation}
	Using the Hadamard and compatibility conditions, see e.g.\ \citet{Chen1973}, \citet{Truesdell1960}, equations \eqref{eq:x-jump}, \eqref{eq:m-jump} and \eqref{eq:eta-jump} may be rearranged as
\begin{align}
	  & (\rho U_N^2\delta_{ij}-Q_{ij})a_j = 0, 
	  	\label{eq:ai1} \\
	- & \left(U_N N_A\frac{\p m}{\p F_{iA}}+
		\frac{\p J_A}{\p F_{iB}}N_A N_B\right)a_i = \notag\\
	  & \qquad\qquad=\frac{\p J_A}{\p p_{,B}}N_A N_B b+
	  	\frac{\p J_A}{\p\theta_{,B}}N_A N_B c, 
		\label{eq:ai2} \\
	  & \left(\rho\theta\frac{\p\eta}{\p F_{iA}}U_N N_A-
	  	\frac{\p q_A}{\p F_{iA}}N_A N_B\right)a_i = \notag\\
	  & \qquad\qquad=\frac{\p q_A}{\p\theta_{,B}}N_A N_B c+
		\frac{\p q_A}{\p p_{,B}}N_A N_B b,
		\label{eq:ai3}
\end{align}
where $Q_{ij}$ is the acoustic tensor, namely
\begin{equation}
	Q_{ij} = \rho\frac{\p^2\psi}{\p F_{jB}\p F_{iA}}N_A N_B,
	\label{eq:acoustic-tensor}
\end{equation}
$U_N$ is the speed of $\mathscr{S}$ at the point $\mathbf{X}$, and $N_A$ is the unit normal to $\mathscr{S}$ at $\mathbf{X}$ in the reference configuration.

	To discuss propagation conditions from \eqref{eq:ai1} we use the relation
\begin{equation}
	N_A = F_{iA}n_i\frac{|\nabla_{\mathbf{x}}\mathfrak{s}|}{\nabla_{\mathbf{X}}\mathscr{S}}
\end{equation}
see \citet{Truesdell1960}, eq.\ (182.8), where $\mathfrak{s}$ and $n_i$ correspond to $\mathscr{S}$ and $N_A$, but in the current configuration.
	One now rewrites $Q_{ij}$ in \eqref{eq:acoustic-tensor} as a function $Q_{ij}(\mathbf{n},U_N)$ in the current configuration to deduce an acceleration wave may propagate provided $a_i$ is an eigenvector of $Q_{ij}$, see \citet{Truesdell1965}, p.271. 
	Existence results for longitudinal and transverse waves are discussed at length in \citet{Truesdell1966} and in \citet{Chen1973}, pp.\ 316--322, and the arguments given there hold also for the case in hand.
	Thus the wavespeed follows from
$$
	\rho U_N^2 = |\mathbf{Q}(\mathbf{N})\mathbf{n}|.
$$

	Once the amplitude $a_i$ is determined equations \eqref{eq:ai2} and \eqref{eq:ai3} become a system of two simultaneous linear equations which yield the pressure and thermal amplitudes $b$ and $c$.

\section{Amplitude calculation}

	We calculate the amplitudes in the case of a one-dimensional wave. 
	It is possible to calculate the amplitudes for a three-dimensional wave but the differential geometry involved is technical and the one-dimensional case yields much of the associated physics.

	The one dimensional equivalents of equations \eqref{eq:motion}, \eqref{eq:energy} and \eqref{eq:m-motion} may be written
\begin{equation}
	\ddot{u} = \frac{\p\psi_F}{\p X},\qquad 
		\dot{m} = \frac{\p J}{\p X},\qquad
		\rho\theta\dot\eta = -\frac{\p q}{\p X},
	\label{eq:one-dim-fields}
\end{equation}
and the associated constitutive theory is
\begin{equation}
	\begin{aligned}
		\psi &= \psi(F,p,\theta),
			& m &= m(F,p,\theta), \\
		 J   &= J(F,p,p_X,\theta,\theta_X),\quad 
		 	& q &= q(F,p,p_X,\theta,\theta_X).
	\end{aligned}
\end{equation}

	We suppose the wave is moving into a region in which $u_X$, $p$ and $\theta$ are constants, so that $u_X^+$, $p^+$ and $\theta^+$ are constants, where the jump notation $[\ddot{u}] = \ddot{u}^- -\ddot{u}^+$ is used. 
	The idea is to differentiate equation \eqref{eq:one-dim-fields}, and take the jumps, and then employ the one-dimensional equivalents of equations \eqref{eq:ai1}, \eqref{eq:ai2} and \eqref{eq:ai3} together with the Hadamard relation and the equation for the jump of a product.
	Since the calculations are now well known we simply state the final result.

	Denote by
$$
	a = [\ddot{u}],\qquad b = [\ddot{p}],\qquad c = [\ddot\theta],
$$
and then one may show
\begin{equation}
	\frac{\delta a}{\delta t}+ka^2-\gamma a = 0,
	\label{eq:bernoulli}
\end{equation}
where $\delta/\delta t$ is the rate of change seen by an observer on the wave, and the coefficients $k$ and $\gamma$ have form
\begin{equation}
	k = \frac{\psi_{FFF}}{2U_N^3},\qquad 
		\gamma = \frac{\alpha}{2U_N}\psi_{F_p}-\frac{\beta}{2U_N}\psi_{F_\theta},
		\label{eq:coeffs}
\end{equation}
where
\begin{equation}
	\begin{aligned}
		\alpha &= \frac1D\{q_{\theta_X}(J_F+U_N m_F)+
			J_{\theta_X}(-q_F+\rho\theta\eta_F U_N)\}, \\ 
		\beta  &= \frac1D\{q_{p_X}(J_F+U_N m_F)+
			J_{p_X}(-q_F+\rho\theta\eta_F U_N)\}, \\
		 D     &= J_{p_X}q_{\theta_X}-J_{\theta_X}q_{p_X}.
	\end{aligned}
		\label{eq:coeffs-params}
\end{equation}

	The solution to \eqref{eq:bernoulli} is
\begin{equation}
	a(t) = \frac{a(0)}{e^{-\gamma t}+(ka(0)/\gamma)[1-e^{-\gamma t}]}.
		\label{eq:a-amplitude}
\end{equation}
	When $a(0)<0$ the wave amplitude $a(t)$ blows-up in a finite time
\begin{equation}
	T = \frac{1}{\gamma}\log\left[\frac{ka(0)-\gamma}{ka(0)}\right].
		\label{eq:blow-up-time}
\end{equation}

	The amplitudes $b$ and $c$ follow from \eqref{eq:a-amplitude} and use of relations
$$
	[p_{XX}] = -\frac{\alpha}{U_N^2}a,\qquad [\theta_{XX}] = \frac{\beta}{U_N^2}a.
$$

	It is worth comparing \eqref{eq:a-amplitude} and \eqref{eq:blow-up-time} to the equivalent expressions in the isothermal case with zero pores, cf.\ \citet{Straughan2008}, p.304, and the isothermal case with pores, cf.\ \citet{Ciarletta2018}.
	For all three cases $k$ has the same value.
	However, $\gamma$ is not present in the isothermal, zero pore case, whereas $\gamma = (m_F U_N+J_F)\psi_{F_p}/2U_N J_{p_X}$ for the isothermal case with pores, see \citet{Ciarletta2018}. 
	The effect of the inclusion of the thermal terms is clearly seen in \eqref{eq:coeffs}, \eqref{eq:coeffs-params} and \eqref{eq:blow-up-time}.

\section{Conclusions}

	We have presented a theory for the evolutionary behaviour of a thermoelastic body which contains pores.
	The theory involves three variables, namely, the displacement $x_i$, the temperature $\theta$, and the pressure in the pores $p$.
	
	A fully nonlinear acceleration wave analysis is performed.
	The wavespeed is calculated for an acceleration wave in the three-dimensional case and this has the same form as that of classical nonlinear thermoelasticity.
	The wave amplitudes are determined for a one-dimensional acceleration wave moving into an equilibrium region.
	In this case $[\ddot{u}] \equiv \ddot{u}^-$ where the minus indicates the value at the left of the wave and if $\ddot{u}^-(\mathbf{X},0) < 0$ it is found that the amplitude may blow-up in a finite time $T$.
	Since $\ddot{u}^- = U_N^2 u_{XX}^-$ this means $u_{XX}^-\to-\infty$ in a finite time which is suggestive of the formation of a shock wave at time $T$.
	
	In the purely isothermal elastic case $T = -1/ka(0)$ whereas $T$ involves $\gamma$ when a porous elastic body is employed.
	The precise effect of temperature and the porosity is given by equation \eqref{eq:blow-up-time} which involves derivatives of the variables $m$, $\eta$, $J$ and $q$.


\begin{thebibliography}{33}
\expandafter\ifx\csname natexlab\endcsname\relax\def\natexlab#1{#1}\fi
\providecommand{\url}[1]{\texttt{#1}}
\providecommand{\href}[2]{#2}
\providecommand{\path}[1]{#1}
\providecommand{\DOIprefix}{doi:}
\providecommand{\ArXivprefix}{arXiv:}
\providecommand{\URLprefix}{URL: }
\providecommand{\Pubmedprefix}{pmid:}
\providecommand{\doi}[1]{\href{http://dx.doi.org/#1}{\path{#1}}}
\providecommand{\Pubmed}[1]{\href{pmid:#1}{\path{#1}}}
\providecommand{\bibinfo}[2]{#2}
\ifx\xfnm\relax \def\xfnm[#1]{\unskip,\space#1}\fi
\bibitem[{Biot(1962)}]{Biot1962}
\bibinfo{author}{M.~Biot},
\newblock \bibinfo{title}{{G}eneralized {T}heory of {A}coustic {P}ropagation in
  {P}orous {D}issipative {M}edia},
\newblock \bibinfo{journal}{J. Acoust. Soc. Am.} \bibinfo{volume}{34}
  (\bibinfo{year}{1962}) \bibinfo{pages}{1254--1264}.
\bibitem[{Brunnhuber and Jordan(2016)}]{Brunnhuber2016}
\bibinfo{author}{R.~Brunnhuber}, \bibinfo{author}{P.~Jordan},
\newblock \bibinfo{title}{{O}n the reduction of {B}lackstock's model of
  thermoviscous compressible flow via {B}ecker's assumption},
\newblock \bibinfo{journal}{Int. J. Non Linear Mech.} \bibinfo{volume}{78}
  (\bibinfo{year}{2016}) \bibinfo{pages}{131--132}.
\bibitem[{Christov(2016)}]{Christov2016a}
\bibinfo{author}{I.~Christov},
\newblock \bibinfo{title}{{N}onlinear acoustics and shock formation in lossless
  barotropic {G}reen--{N}aghdi fluids},
\newblock \bibinfo{journal}{Evol. Equ. Control Theory} \bibinfo{volume}{5}
  (\bibinfo{year}{2016}) \bibinfo{pages}{349--365}.
\bibitem[{Christov and Jordan(2010)}]{Christov2010}
\bibinfo{author}{I.~Christov}, \bibinfo{author}{P.~Jordan},
\newblock \bibinfo{title}{{O}n the propagation of second-sound in nonlinear
  media: {S}hock, acceleration and traveling wave results},
\newblock \bibinfo{journal}{J. Therm. Stresses} \bibinfo{volume}{33}
  (\bibinfo{year}{2010}) \bibinfo{pages}{1109--1135}.
\bibitem[{Christov et~al.(2016)Christov, Jordan, Chin-Bing, and
  Warn-Varnas}]{Christov2016b}
\bibinfo{author}{I.~Christov}, \bibinfo{author}{P.~Jordan},
  \bibinfo{author}{S.~Chin-Bing}, \bibinfo{author}{A.~Warn-Varnas},
\newblock \bibinfo{title}{{A}coustic traveling waves in thermoviscous perfect
  gases: {K}inks, acceleration waves, and shocks under the {T}aylor-{L}ighthill
  balance},
\newblock \bibinfo{journal}{Math. Comput. Simul} \bibinfo{volume}{127}
  (\bibinfo{year}{2016}) \bibinfo{pages}{2--18}.
\bibitem[{Ciarletta and Straughan(2006)}]{Ciarletta2006}
\bibinfo{author}{M.~Ciarletta}, \bibinfo{author}{B.~Straughan},
\newblock \bibinfo{title}{{P}oroacoustic acceleration waves},
\newblock \bibinfo{journal}{Proc. R. Soc. London, Ser. A} \bibinfo{volume}{462}
  (\bibinfo{year}{2006}) \bibinfo{pages}{3493--3499}.
\bibitem[{Ciarletta and Straughan(2007{\natexlab{a}})}]{Ciarletta2007a}
\bibinfo{author}{M.~Ciarletta}, \bibinfo{author}{B.~Straughan},
\newblock \bibinfo{title}{{T}hermo-poroacoustic acceleration waves in elastic
  materials with voids},
\newblock \bibinfo{journal}{J. Math. Anal. Appl.} \bibinfo{volume}{333}
  (\bibinfo{year}{2007}{\natexlab{a}}) \bibinfo{pages}{142--150}.
\bibitem[{Ciarletta and Straughan(2007{\natexlab{b}})}]{Ciarletta2007b}
\bibinfo{author}{M.~Ciarletta}, \bibinfo{author}{B.~Straughan},
\newblock \bibinfo{title}{{P}oroacoustic acceleration waves with second sound},
\newblock \bibinfo{journal}{J. Sound Vib.} \bibinfo{volume}{306}
  (\bibinfo{year}{2007}{\natexlab{b}}) \bibinfo{pages}{725--731}.
\bibitem[{Jordan(2004)}]{Jordan2004}
\bibinfo{author}{P.~Jordan},
\newblock \bibinfo{title}{{A}n analytical study of {K}uznetsov's equation:
  {D}iffusive solitons, shock formation, and solution bifurcation},
\newblock \bibinfo{journal}{Phys. Lett. A} \bibinfo{volume}{326}
  (\bibinfo{year}{2004}) \bibinfo{pages}{77--84}.
\bibitem[{Jordan(2005)}]{Jordan2005a}
\bibinfo{author}{P.~Jordan},
\newblock \bibinfo{title}{{G}rowth and decay of acoustic acceleration waves in
  {D}arcy-type porous media},
\newblock \bibinfo{journal}{Proc. R. Soc. A} \bibinfo{volume}{461}
  (\bibinfo{year}{2005}) \bibinfo{pages}{2749--2766}.
\bibitem[{Jordan(2006)}]{Jordan2006}
\bibinfo{author}{P.~Jordan},
\newblock \bibinfo{title}{{F}inite-amplitude acoustic traveling waves in a
  fluid that saturates a porous medium: {A}cceleration wave formation},
\newblock \bibinfo{journal}{Phys. Lett. A} \bibinfo{volume}{355}
  (\bibinfo{year}{2006}) \bibinfo{pages}{216--221}.
\bibitem[{Jordan(2009)}]{Jordan2009}
\bibinfo{author}{P.~Jordan},
\newblock \bibinfo{title}{{S}ome remarks on nonlinear poroacoustic phenomena},
\newblock \bibinfo{journal}{Math. Comput. Simul} \bibinfo{volume}{80}
  (\bibinfo{year}{2009}) \bibinfo{pages}{202--211}.
\bibitem[{Jordan(2013)}]{Jordan2013}
\bibinfo{author}{P.~Jordan},
\newblock \bibinfo{title}{{A} note on poroacoustic traveling waves under
  {F}orchheimers law},
\newblock \bibinfo{journal}{Phys. Lett. A} \bibinfo{volume}{377}
  (\bibinfo{year}{2013}) \bibinfo{pages}{1350--1357}.
\bibitem[{Jordan(2016)}]{Jordan2016}
\bibinfo{author}{P.~Jordan},
\newblock \bibinfo{title}{{A} survey of weakly-nonlinear acoustic models:
  1910-2009},
\newblock \bibinfo{journal}{Mech. Res. Commun.} \bibinfo{volume}{73}
  (\bibinfo{year}{2016}) \bibinfo{pages}{127--139}.
\bibitem[{Jordan and Puri(2005)}]{Jordan2005b}
\bibinfo{author}{P.~Jordan}, \bibinfo{author}{A.~Puri},
\newblock \bibinfo{title}{{G}rowth/decay of transverse acceleration waves in
  nonlinear elastic media},
\newblock \bibinfo{journal}{Phys. Lett. A} \bibinfo{volume}{341}
  (\bibinfo{year}{2005}) \bibinfo{pages}{427--434}.
\bibitem[{Jordan and Saccomandi(2012)}]{Jordan2012}
\bibinfo{author}{P.~Jordan}, \bibinfo{author}{G.~Saccomandi},
\newblock \bibinfo{title}{Compact acoustic travelling waves in a class of
  fluids with nonlinear material dispersion},
\newblock \bibinfo{journal}{Proceedings of the Royal Society A: Mathematical,
  Physical and Engineering Sciences} \bibinfo{volume}{468}
  (\bibinfo{year}{2012}) \bibinfo{pages}{3441--3457}.
\bibitem[{Jordan et~al.(2017)Jordan, Passarella, and Tibullo}]{Jordan2017}
\bibinfo{author}{P.~Jordan}, \bibinfo{author}{F.~Passarella},
  \bibinfo{author}{V.~Tibullo},
\newblock \bibinfo{title}{Poroacoustic waves under a mixture-theoretic based
  reformulation of the jordan-darcy-cattaneo model},
\newblock \bibinfo{journal}{Wave Motion} \bibinfo{volume}{71}
  (\bibinfo{year}{2017}) \bibinfo{pages}{82--92}.
\bibitem[{Jordan et~al.(2018)Jordan, Keiffer, and Saccomandi}]{Jordan2018}
\bibinfo{author}{P.~Jordan}, \bibinfo{author}{R.~Keiffer},
  \bibinfo{author}{G.~Saccomandi},
\newblock \bibinfo{title}{A re-examination of weakly-nonlinear acoustic
  traveling waves in thermoviscous fluids under rubin-rosenau-gottlieb theory},
\newblock \bibinfo{journal}{Wave Motion} \bibinfo{volume}{76}
  (\bibinfo{year}{2018}) \bibinfo{pages}{1--8}.
\bibitem[{Paoletti(2012)}]{Paoletti2012}
\bibinfo{author}{P.~Paoletti},
\newblock \bibinfo{title}{{A}cceleration waves in complex materials},
\newblock \bibinfo{journal}{Discrete Contin. Dyn. Syst. Ser. B}
  \bibinfo{volume}{17} (\bibinfo{year}{2012}) \bibinfo{pages}{637--659}.
\bibitem[{Rossmanith and Puri(2016{\natexlab{a}})}]{Rossmanith2016a}
\bibinfo{author}{D.~Rossmanith}, \bibinfo{author}{A.~Puri},
\newblock \bibinfo{title}{{R}ecasting a {B}rinkman-based acoustic model as the
  damped {B}urgers equation},
\newblock \bibinfo{journal}{Evol. Equ. Control Theory} \bibinfo{volume}{5}
  (\bibinfo{year}{2016}{\natexlab{a}}) \bibinfo{pages}{463--474}.
\bibitem[{Rossmanith and Puri(2016{\natexlab{b}})}]{Rossmanith2016b}
\bibinfo{author}{D.~Rossmanith}, \bibinfo{author}{A.~Puri},
\newblock \bibinfo{title}{{N}on-linear evolution of a sinusoidal pulse under a
  {B}rinkman-based poroacoustic model},
\newblock \bibinfo{journal}{Int. J. Non Linear Mech.} \bibinfo{volume}{78}
  (\bibinfo{year}{2016}{\natexlab{b}}) \bibinfo{pages}{53--58}.
\bibitem[{Wei and Jordan(2013)}]{Wei2013}
\bibinfo{author}{D.~Wei}, \bibinfo{author}{P.~Jordan},
\newblock \bibinfo{title}{{A} note on acoustic propagation in power-law fluids:
  {C}ompact kinks, mild discontinuities, and a connection to finite-scale
  theory},
\newblock \bibinfo{journal}{Int. J. Non Linear Mech.} \bibinfo{volume}{48}
  (\bibinfo{year}{2013}) \bibinfo{pages}{72--77}.
\bibitem[{Ciarletta et~al.(2018)Ciarletta, Straughan, and
  Tibullo}]{Ciarletta2018}
\bibinfo{author}{M.~Ciarletta}, \bibinfo{author}{B.~Straughan},
  \bibinfo{author}{V.~Tibullo},
\newblock \bibinfo{title}{Acceleration waves in a nonlinear biot theory of
  porous media},
\newblock \bibinfo{journal}{Int. J. Non Linear Mech.}  (\bibinfo{year}{2018}).
  \bibinfo{note}{Submitted}.
\bibitem[{Biot(1972)}]{Biot1972}
\bibinfo{author}{M.~Biot},
\newblock \bibinfo{title}{{T}heory of finite deformations of porous solids},
\newblock \bibinfo{journal}{Indiana Univ. Math. J.} \bibinfo{volume}{21}
  (\bibinfo{year}{1972}) \bibinfo{pages}{597--620}.
\bibitem[{De~Boer and Liu(1995)}]{DeBoer1995}
\bibinfo{author}{R.~De~Boer}, \bibinfo{author}{Z.~Liu},
\newblock \bibinfo{title}{{P}ropagation of acceleration waves in incompressible
  saturated porous solids},
\newblock \bibinfo{journal}{Transp. Porous Media} \bibinfo{volume}{21}
  (\bibinfo{year}{1995}) \bibinfo{pages}{163--173}.
\bibitem[{Iesan(2004)}]{Iesan2004}
\bibinfo{author}{D.~Iesan}, \bibinfo{title}{{T}hermoelastic models of
  continua}, \bibinfo{publisher}{Springer}, \bibinfo{address}{Dordrecht},
  \bibinfo{year}{2004}.
\bibitem[{Straughan(2017)}]{Straughan2017}
\bibinfo{author}{B.~Straughan}, \bibinfo{title}{Mathematical Aspects of
  Multi-Porosity Continua}, volume~\bibinfo{volume}{38} of
  \textit{\bibinfo{series}{Advances in Mechanics and Mathematics}},
  \bibinfo{publisher}{Springer International Publishing}, \bibinfo{year}{2017}.
  \DOIprefix\doi{10.1007/978-3-319-70172-1}.
\bibitem[{Chen(1984)}]{Chen1984}
\bibinfo{author}{P.~Chen},
\newblock \bibinfo{title}{{T}hermodynamic effects in wave propagation},
\newblock in: \bibinfo{booktitle}{Rational Thermodynamics},
  \bibinfo{publisher}{Springer}, \bibinfo{year}{1984}, pp.
  \bibinfo{pages}{191--210}.
\bibitem[{Chen(1973)}]{Chen1973}
\bibinfo{author}{P.~Chen},
\newblock \bibinfo{title}{Growth and decay of waves in solids},
\newblock in: \bibinfo{editor}{C.~Truesdell} (Ed.),
  \bibinfo{booktitle}{Mechanics of Solids}, volume \bibinfo{volume}{VIa/3} of
  \textit{\bibinfo{series}{Encyclopedia of Physics}},
  \bibinfo{publisher}{Springer-Verlag, Berlin Heidelberg New York},
  \bibinfo{year}{1973}, pp. \bibinfo{pages}{303---402}.
\bibitem[{Truesdell and Toupin(1960)}]{Truesdell1960}
\bibinfo{author}{C.~Truesdell}, \bibinfo{author}{R.~Toupin},
\newblock \bibinfo{title}{The classical field theories},
\newblock in: \bibinfo{editor}{S.~Fl\"ugge} (Ed.),
  \bibinfo{booktitle}{Principles of Classical Mechanics and Field Theory},
  volume \bibinfo{volume}{III/1} of \textit{\bibinfo{series}{Encyclopedia of
  Physics}}, \bibinfo{publisher}{Springer-Verlag, Berlin Heidelberg GmbH},
  \bibinfo{year}{1960}, pp. \bibinfo{pages}{226--793}.
\bibitem[{Truesdell and Noll(1965)}]{Truesdell1965}
\bibinfo{author}{C.~Truesdell}, \bibinfo{author}{W.~Noll},
\newblock \bibinfo{title}{The non-linear field theories of mechanics},
\newblock in: \bibinfo{editor}{S.~Fl\"ugge} (Ed.), \bibinfo{booktitle}{The
  Non-Linear Field Theories of Mechanics}, volume \bibinfo{volume}{III/3} of
  \textit{\bibinfo{series}{Encyclopedia of Physics}},
  \bibinfo{publisher}{Springer-Verlag, Berlin, Heidelberg, New York},
  \bibinfo{year}{1965}, pp. \bibinfo{pages}{1--602}.
\bibitem[{Truesdell(1966)}]{Truesdell1966}
\bibinfo{author}{C.~Truesdell},
\newblock \bibinfo{title}{{E}xistence of {L}ongitudinal {W}aves},
\newblock \bibinfo{journal}{J. Acoust. Soc. Am.} \bibinfo{volume}{40}
  (\bibinfo{year}{1966}) \bibinfo{pages}{729--730}.
\bibitem[{Straughan(2008)}]{Straughan2008}
\bibinfo{author}{B.~Straughan}, \bibinfo{title}{Stability and Wave Motion in
  Porous Media}, volume \bibinfo{volume}{165} of
  \textit{\bibinfo{series}{Advances in Mechanics and Mathematics}},
  \bibinfo{publisher}{Springer-Verlag New York}, \bibinfo{year}{2008}.

\end{thebibliography}
\end{document}